\documentclass[conference]{IEEEtran}
\IEEEoverridecommandlockouts
\usepackage{cite}
\usepackage{amsmath,amssymb,amsfonts}
\usepackage{algorithmic}
\usepackage{graphicx}
\usepackage{textcomp}
\usepackage{xcolor}
\def\BibTeX{{\rm B\kern-.05em{\sc i\kern-.025em b}\kern-.08em
    T\kern-.1667em\lower.7ex\hbox{E}\kern-.125emX}}
\begin{document}

\title{Non Parametric Data Augmentations Improve Deep-Learning based Brain Tumor Segmentation\\
}

\author{\IEEEauthorblockN{Hadas Ben Atya, Ori Rajchert}
\IEEEauthorblockA{\textit{Faculty of Biomedical Engineering} \\
\textit{Technion - IIT}\\
 Haifa, Israel \\
\{hds,ori.ra\}@campus.technion.ac.il}
\and
\IEEEauthorblockN{Liran Goshen}
\IEEEauthorblockA{\textit{Global Advanced Technology, CT BU} \\
\textit{Philips}\\
 Haifa, Israel \\
liran.goshen@philips.com}
\and
\IEEEauthorblockN{Moti Freiman}
\IEEEauthorblockA{\textit{Faculty of Biomedical Engineering} \\
\textit{Technion - IIT}\\
 Haifa, Israel \\
moti.freiman@technion.ac.il}}

\maketitle

\begin{abstract}
Automatic brain tumor segmentation from Magnetic Resonance Imaging (MRI) data plays an important role in assessing tumor response to therapy and personalized treatment stratification. Manual segmentation is tedious and subjective. Deep-learning based algorithms for brain tumor segmentation have the potential to provide objective and fast tumor segmentation. However, the training of such algorithms requires large datasets which are not always available. Data augmentation techniques may reduce the need for large datasets. However current approaches are mostly parametric and may  result in suboptimal performance. We introduce two non-parametric methods of data augmentation for brain tumor segmentation: the mixed structure regularization (MSR) and shuffle pixels noise (SPN). We evaluated the added value of the MSR and SPN augmentation on the brain tumor segmentation (BraTS) 2018 challenge dataset with the encoder-decoder nnU-Net architecture as the segmentation algorithm. Both MSR ans SPN improve the nnU-Net segmentation accuracy compared to parametric Gaussian noise augmentation.(Mean dice score increased from 80\% to 82\% and p-values=0.0022, 0.0028 when comparing MSR to non parametric augmentation for the tumor core and whole tumor experiments respectively. The proposed MSR and SPN augmentations has the potential to improve neural-networks performance in other tasks as well. 
\end{abstract}


\section{Introduction}
Primary brain tumors originate from brain cells while secondary tumors stem from other organs. Brain tumors are usually discovered by  Magnetic Resonance Imaging (MRI) data including T1-weighted, Contrast-enhanced T1-weighted, T2-weighted, and Fluid Attenuation Inversion Recovery (FLAIR) sequences \cite{b_1_milletari2016v}.
 Early diagnosis of brain tumors plays an important role in improving treatment possibilities and increases the survival rate of the patients \cite{b_2_wang2019automatic}. 
 Segmentation of the tumor and its subregions from the MRI data is necessary for tumor initial grading, response to therapy assessment and personalized treatment stratification.
 
 Manual segmentation of the tumor from the multi-modal MRI data by radiologist/technologist is time-consuming and subjective  \cite{b_3_myronenko20183d}. In contrast automatic segmentation has the potential to provide objective and reproducible measurements of the tumor. Further, automatic methods can integrate information from the various MRI sequences rather than relying upon a single sequence. Therefore, it can significantly improve analysis accuracy, quality of care, and treatment by supporting optimal therapy, surgical planning, and monitoring.

Neural networks are the state of the art for automatic segmentation in many fields, especially the encoder-decoder architectures which were first introduced by the U-Net \cite{b_4_ronneberger2015u}, and recently improved by the nnU-Net \cite{b_5_isensee2019automated}.
However, these methods require large amounts of data which are not always available for medical imaging tasks. A common approach to overcome this obstacle, is to apply data augmentation techniques such as random Gaussian noise to increase the variety of the training dataset and improve the robustness of the model.

However, adding such parametric noise may not enlarge the data variety as the neural networks may learn how to denoise it rather than improve its segmentation capabilities.

In this work we introduce the  mixed structure regularization (MSR) in \cite{b_7_freiman2019unsupervised}, and shuffle pixels noise (SPN) for brain tumor segmentation. These augmentations are non-parametric and thus have the potential to encourage the network to generalize better by increasing the variety of the training dataset rather than learning to denoise the data. Therefore, the methods have the potential to improve the robustness of our brain tumor segmentation model.

We demonstrated the added value of the MSR and SPN augmentation on the brain tumor segmentation (BraTS) 2018 challenge dataset \cite{b_9_menze2014multimodal} with the encoder-decoder nnU-Net architecture as the segmentation algorithm. Both MSR ans SPN improve the nnU-Net segmentation accuracy compared to parametric Gaussian noise augmentation. Our results suggest that MSR and SPN can improve neural networks accuracy and reduce the need for large databases, especially for medical image analysis tasks.

\section{Method}

\subsection{CNN based segmentation}
We trained our model with the dynamic U-Net (DynUNet) neural network implemented with Pytorch and MONAI, which is based on the nnU-Net described in \cite{b_5_isensee2019automated}.   

The DynUNet is based on a 3D U-Net encoder-decoder model, with five downsampling operations and five upsampaling operations. Downsampling is performed with strided convolutions, and upsampling is performed with transposed convolution as described in \cite{b_8_isensee2021nnu}

The model used Leaky ReLU nonlinearities activation function and batch normalization for feature map normalization.
We trained the model for 300 epochs, with an input patch size of 128X128X128 and a batch size equal to 5.

A stochastic gradient descent based pyTorch was used with an initial learning rate of 0.01 and Nesterov momentum of 0.99. The learning rate was decreased by a polynomial schedule:   
\begin{equation}
\label{eq:polyDecyed}
lr= lr_0 * (1- \frac{epoch}{epoch_{max}})^{0.9}
\end{equation}

A gradient-based learning optimization process is challenging due to gradient vanishing effects, which make the loss back-propagation ineffective in early layers \cite{b_6_dou20163d}. 
To overcome this challenge, an auxiliary segmentation output is added to all layers but the three lowest resolutions in the decoder, by applying a 1x1x1 convolution with a softmax layer to obtain dense predictions and calculate its loss \cite{b_5_isensee2019automated, b_6_dou20163d}.

Our loss function for this model is a weighted sum of the losses at all resolutions:
\begin{equation}
\label{eq:loss}
L=w_0*L_0+w_1*L_1+ w_2*L_2
\end{equation}

Where $w_0$ corresponds to the highest resolution, and the weights decrease with each decrease in the resolution as a function of $w_i=\frac{1}{2}^i$ \cite{b_5_isensee2019automated}.

For the loss function, we used a combination of dice loss and binary cross-entropy (BCE) as described in \cite{b_8_isensee2021nnu}. 
BCE is a loss function that is used in binary classification tasks:
\begin{equation}
BCE = -\frac{1}{n}\sum_{i=1}^N{y_i\log{(\hat{y_i})}+{(1-y_i)}\log{(1-\hat{y_i})}}
\end{equation}
Where $y_i$ is the target label $\hat{y_i}$ is the predicted label, and N is the output size.

These tasks answer a question with only two choices (yes or no, A or B, 0 or 1, left or right). Formally, this loss is equal to the average of the categorical cross-entropy (CE) loss on many two-category tasks:
\begin{equation}
CE = -\sum_{i=1}^N{y_i\log{(\hat{y_i})}}
\end{equation}

\subsection{Data augmentation}

In order to increase the variety of our training dataset and increase the robustness of our model, data augmentations were applied. 
Our baseline transformations were implemented with MONAI and included:
\begin{itemize}
\item Normalize the image intensity using calculated mean and std on each channel separately to all non-zero voxels.
\item Randomly scale the intensity of the input images with a probability of 0.3.
\item Randomly shift the intensity with an offset range generated from Union[-0.1,0.1] and a probability of shift equals 0.3.
\item Random spatial cropping of the images and the target label with an ROI size of [128,128,128].
\item Random flipping of the images and labels along the z axes with probability of 0.3 of flipping.
\item Random elastic deformation and a 3D affine transformation with a probability of 0.3. The random offsets on the grid were generated from a Uniform[0.1,0.3], as well as the standard deviation of a Gaussian kernel which was used to smooth the random offset grid and the shear range.  
\end{itemize}

We compared the baseline model with other models which included those transforms with an addition of MSR, SPN or Gaussian-noise separately. 

\subsection{Mixed structure regularization (MSR):}
We corrupted the input data by randomly adding different brain images from our training set, as described by \cite{b_7_freiman2019unsupervised}. 
We can represent this data augmentation with the function:
\begin{equation}
MSR(x)=(1-\alpha)*x+\alpha*x_r
\end{equation}
Where x is the original image, and $x_r$ is a randomly selected image. 

We applied this augmentation multiple times with a different magnitude $= \alpha$  and different probability. 
We saw that the best probability and magnitude for our dataset is: $p=0.5, \alpha=1*10^-4$.

\begin{figure}[htbp]
\centering
\resizebox{0.9\columnwidth}{!}{\includegraphics{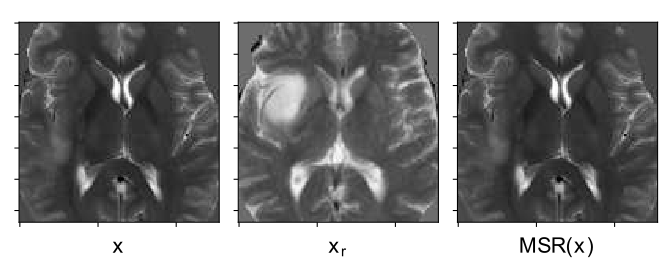}}
\label{fig:MSR}
\caption{An example of the MSR augmentation applied on one of the images from the database.}
\end{figure}

\subsection{Shuffle pixels noise (SPN):}
A random permutation of the pixels is chosen and then the same permutation is applied to all the images in the training set \cite{b_10_zhang2021understanding}. We applied this augmentation multiple times with a different magnitude $= \alpha$.
\begin{equation}
SPN(x)=(1-\alpha)*x+\alpha*x_r
\end{equation}
Where x is the original image, and $x_r$ is the image after shuffling pixels on the x,y axis.
We saw that the best probability and magnitude for our dataset is: $p=1, \alpha=1*10^-7$.
    \begin{figure}[htbp]
    \centering
    \resizebox{0.9\columnwidth}{!}{\includegraphics{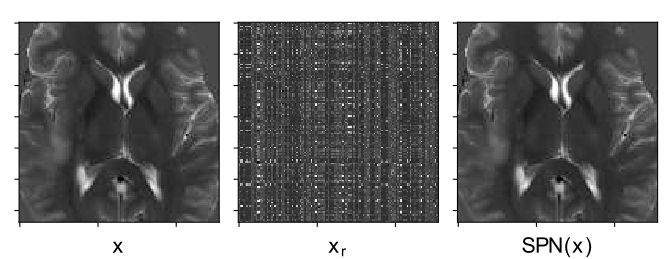}}
    \label{fig:SPN augmentation}
    \caption{An example of the SPN augmentation applied on one of the images from the database.}
    \end{figure}

\section{Results}
\subsection{BraTS database:}\label{AA}
We used the BraTS 2018 challenge dataset, which contains 285 multimodal scans - a) native (T1) and b) post-contrast T1-weighted (T1Gd), c) T2-weighted (T2), and d) T2 FLAIR volumes.
All images were manually segmented by one to four raters and the segmentations were approved by experienced neuro-radiologists\cite{b_9_menze2014multimodal}.

We randomly split our database into a training, validation, and testing set (80\%,10\%,10\% respectively), while maintaining a ratio between the high-grade gliomas(HGG) and low-grade (LGG) tumors according to the existing ratio in the entire dataset. Our training set contains 229 images, and the validation and testing sets contain 28 images each. 

\begin{figure}[htbp]
\centering
\resizebox{0.7\columnwidth}{!}{\includegraphics{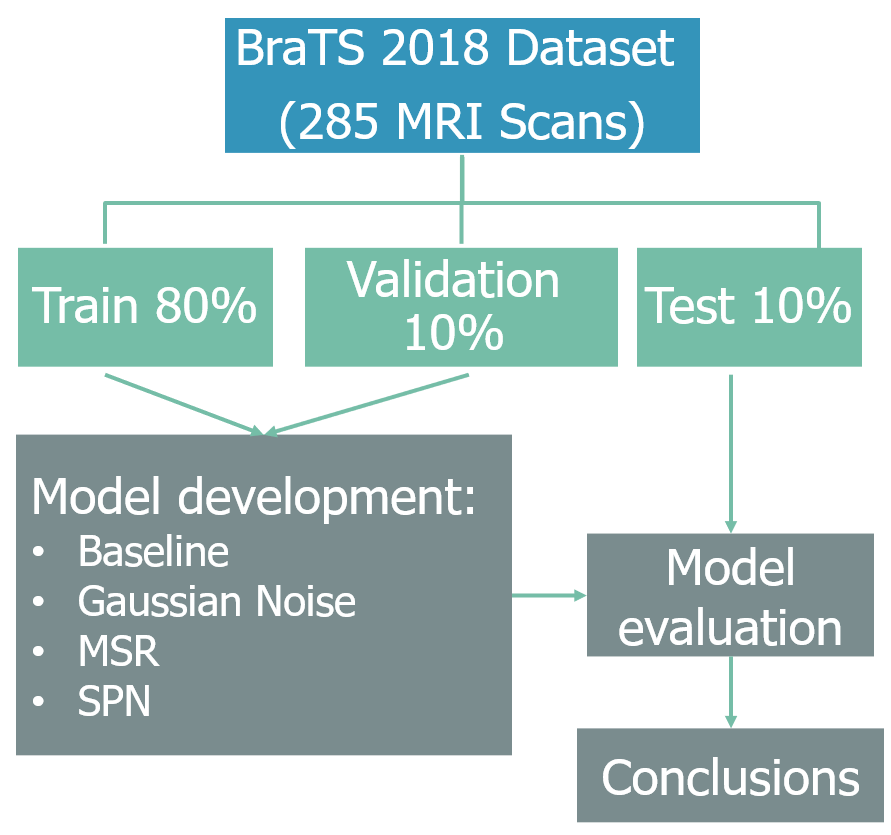}}
\label{fig:Dataset}
\caption{study - flow chart}
\end{figure}

\begin{figure}[htbp]
\centerline{\includegraphics[width=.55\textwidth]{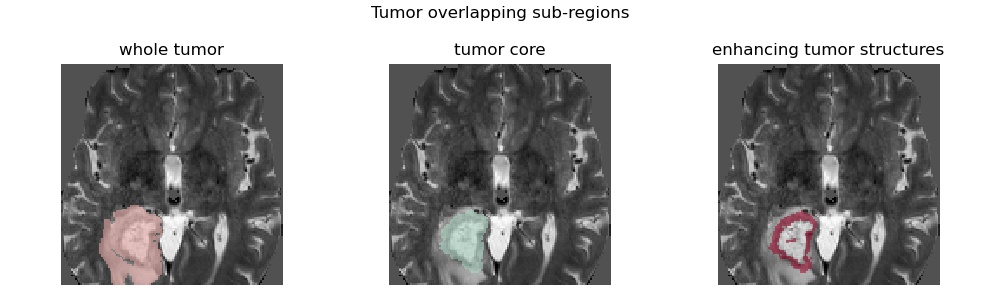}}
\caption{Tumor partially overlapping regions segmentation.}
\label{fig}
\end{figure}

The tumor's regions are `edema', `non-enhancing tumor and necrosis', and `enhancing tumor'. However, we segment three partially overlapping regions, as performed in the BraTS challenge and described in \cite{b_8_isensee2021nnu}: whole tumor (WT) – consisting of all 3 regions; tumor core (TC) – consisting of the non-enhancing tumor and necrosis, and the enhancing tumor (ET) as shown in ``Fig.~\ref{fig}.''
The dice score was then calculated on the validation set for each label separately. Also, we calculated the mean dice of the 3 labels.

\subsection{Test results}
We evaluated our different models on the same test set which contains 27 GBM MRI images. 

\begin{table}[htbp]
\caption{Testing Dataset Results}
\begin{center}
\begin{tabular}{|c|c|c|c|c|}
\hline
\textbf{Augmentation}&\multicolumn{4}{|c|}{\textbf{Dice Mean Score}} \\
\cline{2-5} 
\textbf{Type} & \textbf{\textit{WT}}& \textbf{\textit{TC}}& \textbf{\textit{ET}}& \textbf{\textit{Mean}}\\
\hline
Baseline& $0.80959$& $0.68795$ & $0.78975$ & $0.88811$\\ \hline
Gaussian noise& $0.80775$& $0.69871$& $0.77113$& $0.8918$\\ \hline
SPN& $0.81987$& ${\bf 0.70344}$& $0.79351$& $0.89983$ \\ \hline
\textbf{MSR}& ${\bf0.82396}$ & $0.69705$& ${\bf 0.80732}$ & ${\bf 0.90404}$\\ \hline
\end{tabular}
\label{tab1}
\end{center}
\end{table}


\begin{figure}[htbp]
\centering
\resizebox{\columnwidth}{!}{\includegraphics{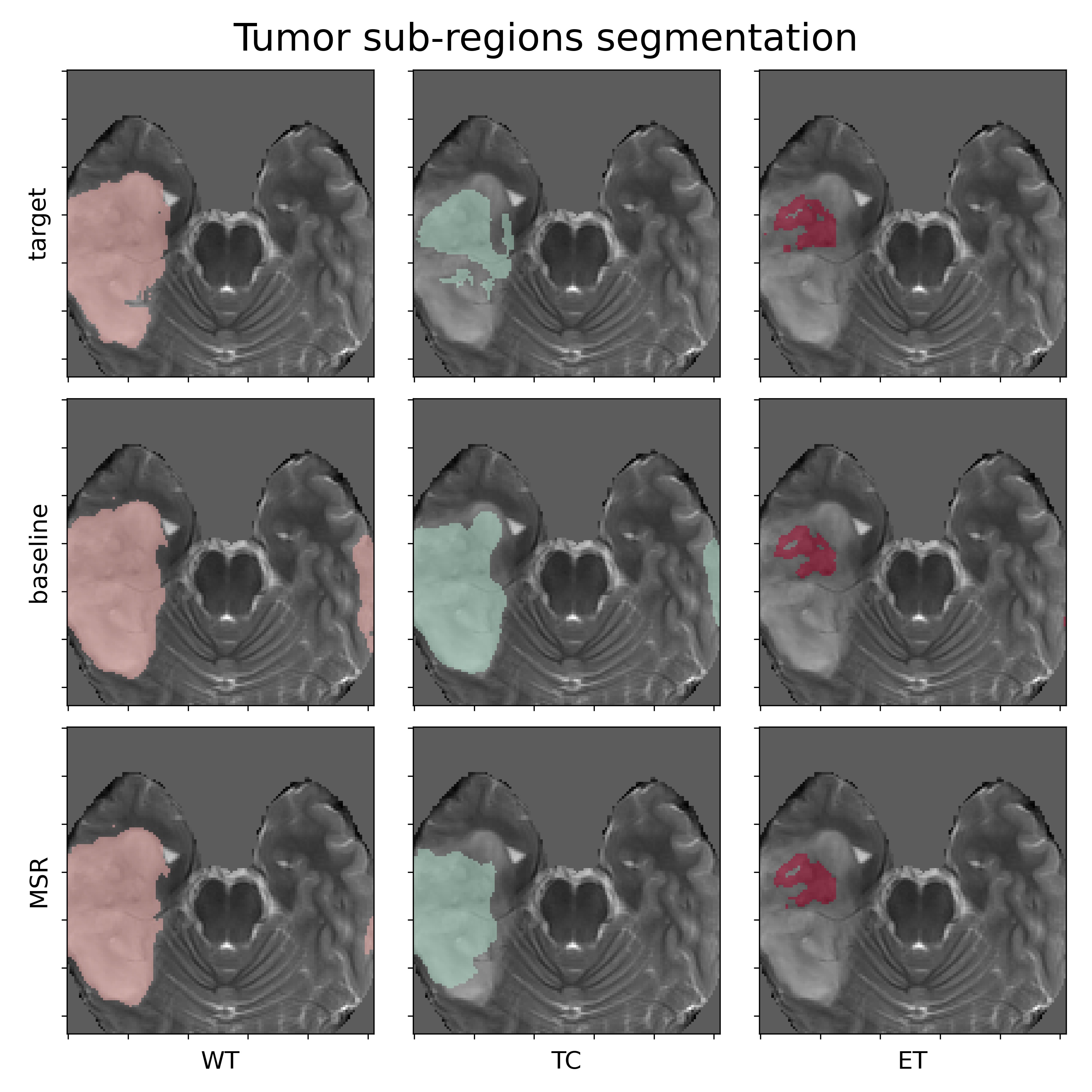}}
\label{fig:GBM}
\caption{Tumor segmentation with nnU-Net. Target structures are shown in 2D (first row), the baseline model's outputs (second row), and our model with mixed structure regularization - MSR outputs (third row).
The MSR reduced over segmentation and achieved better accuracy.}
\end{figure}

\subsection{Statistical analysis}
Comparisons were done with repeated measures ANOVA statistical tests. Paired comparisons were made between the MSR, non parametric augmentation, and the basic parametric augmentation.
These statistical tests showed a significant advantage to MSR (p=0.0022, 0.0028 for the TC, WT respectively).
\newline We visualize our data using box plot to analyze the distribution of our augmentations.
 The whole tumor dice box plot shows that the variance for SPN and MSR is smaller than the other augmentations variance.
A smaller variance indicates higher confidence level of the neural network results.

\begin{figure}[h]
  \includegraphics[width=.97\columnwidth]
    {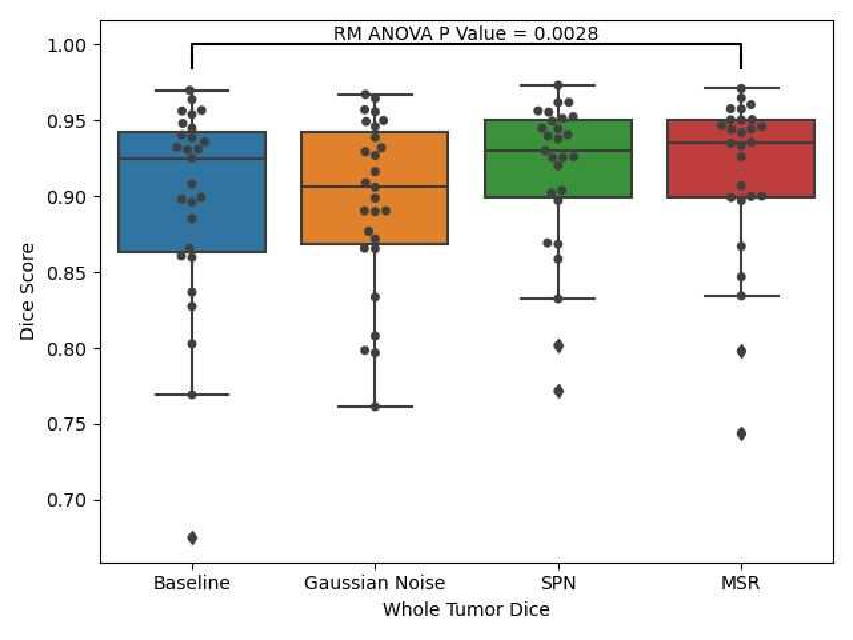}\hfill \\
  \includegraphics[width=.97\columnwidth]
    {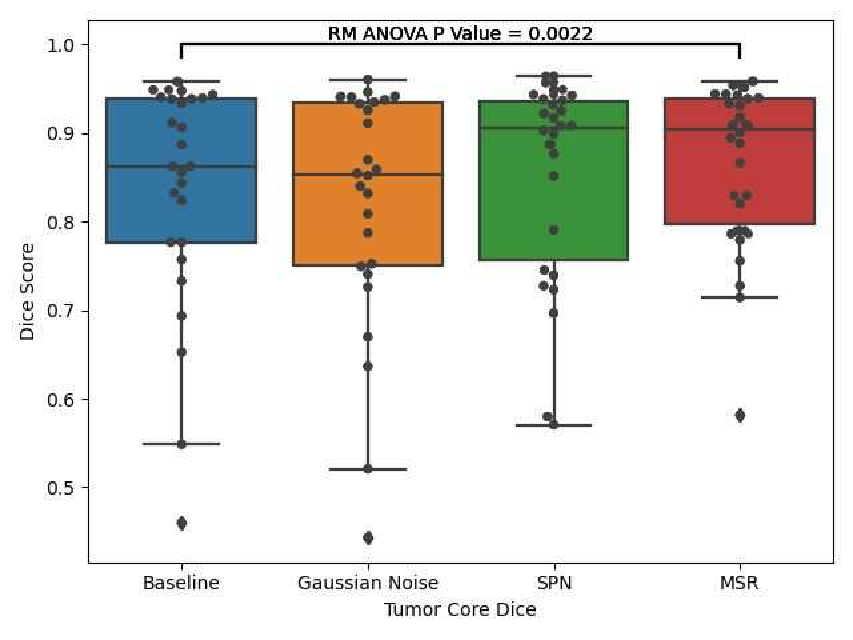}

  \caption{Top: Whole tumor dice multi augmentation box plot. Bottom: Tumor core dice multi augmentation box plot
}
\end{figure}

\subsection{Results}
We saw an improvement in the dice score when using the MSR and the SPN non parametric augmentations. 
By using parametric or Gaussian augmentations the overall dice score achieved was about 80\% Meaning that if two doctors examine a patient's MRI scanned image, both will agree that there is a tumor at a confidence level of 80\%. 
By using the MSR or SPN augmentation we improved the latter figure to about 82\% of confidence level. 
The classification edges that were previously mistakenly suspected to be a tumor are now properly classified as normal tissue.
The augmentations we applied in order to enhance the segmentation quality achieved a statistically significant improvement. We can conclude that these non-parametric models are more accurate.

\subsection{Discussion}
We improved the capacity of the nnU-net to segment brain tumors by introducing the non-parametric MSR and SPN data augmentation techniques. Automatic segmentation for brain tumors from multi-modality MRI data plays a critical role in assessing tumor aggressiveness and response to therapy and in stratifying treatment protocols. While deep-neural-networks have the potential to provide accurate, reliable, and objective segmentation, the training of such algorithms requires large data-sets which are not always available in the medical imaging domain. The non-parametric MSR and SPN data augmentation techniques demonstrated a statistically significant improvement in the accuracy of nnU-net for brain tumor segmentation from multi-modality MRI data using the publicly available Brats 2018 challenge database. The proposed non-parametric augmentations may be further extended to achieve additional improvements in deep-neural-networks algorithms for various biomedical imaging tasks.





\bibliographystyle{IEEEtran}
\bibliography{conference_041818}

\begin{thebibliography}{10}
\providecommand{\url}[1]{#1}
\csname url@samestyle\endcsname
\providecommand{\newblock}{\relax}
\providecommand{\bibinfo}[2]{#2}
\providecommand{\BIBentrySTDinterwordspacing}{\spaceskip=0pt\relax}
\providecommand{\BIBentryALTinterwordstretchfactor}{4}
\providecommand{\BIBentryALTinterwordspacing}{\spaceskip=\fontdimen2\font plus
\BIBentryALTinterwordstretchfactor\fontdimen3\font minus
  \fontdimen4\font\relax}
\providecommand{\BIBforeignlanguage}[2]{{%
\expandafter\ifx\csname l@#1\endcsname\relax
\typeout{** WARNING: IEEEtran.bst: No hyphenation pattern has been}%
\typeout{** loaded for the language `#1'. Using the pattern for}%
\typeout{** the default language instead.}%
\else
\language=\csname l@#1\endcsname
\fi
#2}}
\providecommand{\BIBdecl}{\relax}
\BIBdecl

\bibitem{b_1_milletari2016v}
F.~Milletari, N.~Navab, and S.-A. Ahmadi, ``V-net: Fully convolutional neural
  networks for volumetric medical image segmentation,'' in \emph{2016 fourth
  international conference on 3D vision (3DV)}.\hskip 1em plus 0.5em minus
  0.4em\relax IEEE, 2016, pp. 565--571.

\bibitem{b_2_wang2019automatic}
G.~Wang, W.~Li, S.~Ourselin, and T.~Vercauteren, ``Automatic brain tumor
  segmentation based on cascaded convolutional neural networks with uncertainty
  estimation,'' \emph{Frontiers in computational neuroscience}, vol.~13, p.~56,
  2019.

\bibitem{b_3_myronenko20183d}
A.~Myronenko, ``3d mri brain tumor segmentation using autoencoder
  regularization,'' in \emph{International MICCAI Brainlesion Workshop}.\hskip
  1em plus 0.5em minus 0.4em\relax Springer, 2018, pp. 311--320.

\bibitem{b_4_ronneberger2015u}
O.~Ronneberger, P.~Fischer, and T.~Brox, ``U-net: Convolutional networks for
  biomedical image segmentation,'' in \emph{International Conference on Medical
  image computing and computer-assisted intervention}.\hskip 1em plus 0.5em
  minus 0.4em\relax Springer, 2015, pp. 234--241.

\bibitem{b_5_isensee2019automated}
F.~Isensee, P.~F. J{\"a}ger, S.~A. Kohl, J.~Petersen, and K.~H. Maier-Hein,
  ``Automated design of deep learning methods for biomedical image
  segmentation,'' \emph{arXiv preprint arXiv:1904.08128}, 2019.

\bibitem{b_7_freiman2019unsupervised}
M.~Freiman, R.~Manjeshwar, and L.~Goshen, ``Unsupervised abnormality detection
  through mixed structure regularization (msr) in deep sparse autoencoders,''
  \emph{Medical physics}, vol.~46, no.~5, pp. 2223--2231, 2019.

\bibitem{b_9_menze2014multimodal}
B.~H. Menze, A.~Jakab, S.~Bauer, J.~Kalpathy-Cramer, K.~Farahani, J.~Kirby,
  Y.~Burren, N.~Porz, J.~Slotboom, R.~Wiest \emph{et~al.}, ``The multimodal
  brain tumor image segmentation benchmark (brats),'' \emph{IEEE transactions
  on medical imaging}, vol.~34, no.~10, pp. 1993--2024, 2014.

\bibitem{b_8_isensee2021nnu}
F.~Isensee and K.~H. Maier-Hein, ``nnu-net for brain tumor segmentation,'' in
  \emph{Brainlesion: Glioma, Multiple Sclerosis, Stroke and Traumatic Brain
  Injuries: 6th International Workshop, BrainLes 2020, Held in Conjunction with
  MICCAI 2020, Lima, Peru, October 4, 2020, Revised Selected Papers, Part II},
  vol. 12658.\hskip 1em plus 0.5em minus 0.4em\relax Springer Nature, 2021, p.
  118.

\bibitem{b_6_dou20163d}
Q.~Dou, H.~Chen, Y.~Jin, L.~Yu, J.~Qin, and P.-A. Heng, ``3d deeply supervised
  network for automatic liver segmentation from ct volumes,'' in
  \emph{International conference on medical image computing and
  computer-assisted intervention}.\hskip 1em plus 0.5em minus 0.4em\relax
  Springer, 2016, pp. 149--157.

\bibitem{b_10_zhang2021understanding}
C.~Zhang, S.~Bengio, M.~Hardt, B.~Recht, and O.~Vinyals, ``Understanding deep
  learning (still) requires rethinking generalization,'' \emph{Communications
  of the ACM}, vol.~64, no.~3, pp. 107--115, 2021.

\end{thebibliography}

\end{document}